\documentclass[]{spie}  %>>> use for US letter paper
\usepackage[]{graphicx}

\title{Connecting the time domain community with the Virtual Astronomical Observatory} 

\author{Matthew J. Graham\supit{a},  S.G. Djorgovski\supit{a}, Ciro Donalek\supit{a}, Andrew J. Drake\supit{a}, Ashish A. Mahabal\supit{a}, Raymond L. Plante\supit{b}, Jeffrey Kantor\supit{c} and John C. Good\supit{d}
\skiplinehalf
\supit{a}California Institute of Technology, 1200 E California Blvd, Pasadena, CA 91125, USA; \\
\supit{b}National Center for Supercomputing Applications, University of Illinois Urbana-Champaign, 1205 W. Clark St, Urbana, IL 61801, USA;
\supit{c} LSST Corporation, 950 N Cherry Ave, M-18, Tucson, AZ 85719, USA; 
\supit{d} Infrared Processing and Analysis Center, California Institute of Technology, 770 South Wilson Ave., Pasadena, CA 91125, USA
}

\authorinfo{Further author information: (Send correspondence to M.J.Graham)\\M.J.G.: E-mail: mjg@caltech.edu, Telephone: 1 626 395 8030}
\begin{document} 
  \maketitle 

%%%%%%%%%%%%%%%%%%%%%%%%%%%%%%%%%%%%%%%%%%%%%%%%%%%%%%%%%%%%% 
\begin{abstract}
The time domain has been identified as one of the most important areas of astronomical research for the next decade. The Virtual Observatory is in the vanguard with dedicated tools and services that enable and facilitate the discovery, dissemination and analysis of time domain data. These range in scope from rapid notifications of time-critical astronomical transients to annotating long-term variables with the latest modelling results. In this paper, we will review the prior art in these areas and focus on the capabilities that the VAO is bringing to bear in support of time domain science. In particular, we will focus on the issues involved with the heterogeneous collections of (ancilllary) data associated with astronomical transients, and the time series characterization and classification tools required by the next generation of sky surveys, such as LSST and SKA.

\end{abstract}

%>>>> Include a list of keywords after the abstract 


%%%%%%%%%%%%%%%%%%%%%%%%%%%%%%%%%%%%%%%%%%%%%%%%%%%%%%%%%%%%%
\section{Introduction}
\label{intro}

The time domain is the emerging field of astronomical research, as recognized in the 2010 National Research Councils Decadal Survey of Astronomy and Astrophysics \cite{decadal}. Planned facilities for the next decade and beyond, such as the Large Synoptic Survey Telescope (LSST)\footnote{http://www.lsst.org} and the Square Kilometer Array\footnote{http://www.skatelescope.org} (SKA), will revolutionize our understanding of the universe with nightly searches of large swathes of sky for changing objects and networks of robotic telescopes ready to follow up in greater detail selected interesting sources. This will impact essentially every area of astronomy, from the Solar System to cosmology, and from stellar evolution to extreme relativistic phenomena\cite{ref8}, making it a very rich area for scientific exploration and discovery. Moreover, many interesting phenomena, e.g., supernovae and other types of cosmic explosions, can only be studied in the time domain. 

These new surveys build on a legacy of over fifty years of experience of sky surveys, first with photographic plates and then, more recently,  digital detectors (see Ref.~\citenum{djorgovski} for a recent review). The rise of information technology has driven an exponential growth of data volumes (and, equally importantly, data complexity and data quality) following Moore's law, e.g., DPOSS \cite{ref3} to 2MASS \cite{ref4} to SDSS\cite{ref2}, and many digital sky surveys that followed. To cope with such (necessarily distributed) giga- and terascale data collections,  the community developed the concept of the {\em Virtual Observatory} (VO), which provides the wherewithal to aggregate and analyze disparate data sets, opening up new avenues of scientific research based on data discovery and fusion\cite{vo1, vo2}. The Virtual Astronomical Observatory\cite{vao} (VAO) is the US national VO project and provides the components, libraries, and templates that allow national facilities, major projects, and end-users to craft their own VO-enabled applications for seamless data access and integration, especially in support of data intensive research.

The time domain has been identified early as a prime arena for VO applications\cite{vo3}. It adds a new dimension to data discovery and federation with (near) real-time massive data streams - for example, Palomar-Quest\cite{pq}, Catalina Real-time Transient Survey\cite{crts} (CRTS), Palomar Transient Factory\cite{ptf} (PTF) and Panoramic Survey Telescope And Rapid Response System\cite{ps1}  (Pan-STARRS; PS1) and LSST to come - replacing static data sets; in a way, we've moved from panoramic digital photography of the sky to panoramic digital cinematography. Since many of the observed phenomena in this domain are short-lived, and since the scientific returns depend strongly not only on their detection, but also on the timely and well-chosen follow-up observations, there is a need to fully process the data as they stream from the telescopes, compare it with the previous images of the same parts of the sky, automatically and reliably detect any changes, and classify and prioritize the detected events for the rapid follow-up observations.  This poses significant new technological challenges for the VO and its infrastructure.  Analogous situations may also be found in many other areas, where the data come continuously from some instruments or sensor networks, and where anomalous or specifically targeted events have to be found and responded to in a rapid fashion.  

The VO has evolved a two-track approach to the time domain: one deals specifically with the mechanics of reporting transient celestial events (VOEvent) in a timely fashion and the associated infrastructure to publish, disseminate and archive them. The other deals with the more general issues of time series data, such as how to describe, represent and access them in a way to ensure interoperability between different data archives. In this paper, we will review the specific issues that are associated with the latter and how the VO, and, specifically, the VAO, which is leading the time domain effort, is meeting them. This draws in associated but separate work on source characterization and classification that is an essential part of a time domain system. Details of the transient approach are presented in a complementary paper in these proceedings \cite{skyalert-spie}, although we will discuss certain common issues. Nevertheless, whichever approach is being discussed, operational concerns are an important consideration and we focus particularly on those related to questions of scalability and managing large collections of heterogeneous data.

\section{Interoperable time series}

The promise of data federation is that it can often lead to potentially new scientific insights. An obvious example is combining observations of the same objects from different wavelength regimes, e.g., X-ray, infrared, and radio, to understand the various physical processes that contribute to their spectral energy distributions. The time domain adds an extra dimension to this, allowing the identification of temporally correlated behavior, for example, a X-ray burst followed by an infrared burst  may indicate the propagation of a shock front from an originating source to circumscribing material.

The International Virtual Observatory Alliance\footnote{http://www.ivoa.net} (IVOA) has defined a common set of data access protocols to ensure that the same interface is employed across all data archives, no matter where they are located, to perform the same type of data query (see Table~\ref{table2} for a summary of those defined). 
Although common data formats may be employed in transferring data, e.g., VOTable \cite{votable} for tabular data, individual data providers usually represent and store their data and metadata in their own way. Common data models define the shared elements across data and metadata collections and provide a framework for describing relationships between them so that different representations can interoperate in a transparent manner. 

\begin{table}
\caption{Different types of data access protocol defined by the IVOA.}
\label{table2} 
\begin{tabular}{ll}
\hline\noalign{\smallskip}
Name & Description \\
\noalign{\smallskip}\hline\noalign{\smallskip}
Simple Cone Search (SCS) & Retrieve all objects within a circular region on the sky \\
Simple Image Access (SIA) & Retrieve all images of objects within a region on the sky \\
Simple Spectral Access (SSA) & Retrieve all spectra of objects within a region on the sky \\
Simple Line Access (SLA) & Retrieve spectral line data \\
Simulations (SIMDAL) & Retrieve simulation data \\
Table Access (TAP) & Retrieve tabular data \\
\noalign{\smallskip}\hline
\end{tabular}
\end{table}

VOEvent\cite{voevent} may be regarded as the (lightweight) data model for observations of transient astronomical events, describing their who, what, where/when, how, and why characteristics. However, when individual measurements of arbitrarily named quantities are reported, either as a group of parameters or in a table, their broader context within a standard data model can be established through the IVOA Utypes mechanism \cite{utypes}. These strings act as reference pointers to individual elements within a data model thus identifying the concept that the reported value represents, e.g., the UType ``Data.FluxAxis.Accuracy.StatErrHigh'' identifies a quantity as the upper error bound on a flux value defined in the Spectral data model. Namespaces allow reuse of quantities/concepts defined in one data model in another. 

The Spectral data model \cite{spectraldm} (now in the final stages of specification) defines a generalized model for spectro-photometric sequences and provides a basis for a set of specific case models, such as Spectrum, SED and TimeSeries. The TimeSeries data model is intended to describe any observed or derived quantity that may vary with time with a light curve being the most common example and considered to be a time series with just one photometric band. More complicated time series might involve multi-band data or data where the time sample bin size varies between successive samples. The data model can also describe various levels of associated metadata such as period for the time series if relevant or known and whether the time axis values are folded by the period if one exists, or a target variability amplitude and derived signal-to-noise ratio. A variety of serialization formats for compliant data sets are supported, e.g., FITS, VOTable, and CSV.

\begin{figure}
\includegraphics[width=6.78in]{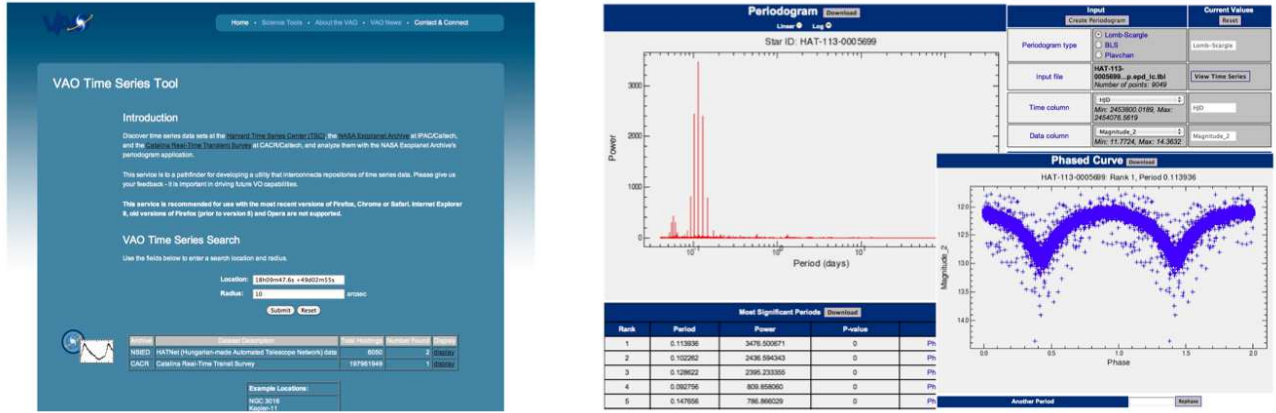}
\caption{This shows a screenshot from the VAO Time Series Search Tool (left) and the Lomb-Scargle periodogram (right) and phased light curve (inset).}
\label{fig1}       % Give a unique label
\end{figure}

A number of data archives are set to expose their time series holdings via the TimeSeries data model. Prime amongst these is the Catalina Real-time Transient Survey\cite{crts} (CRTS) which has light curves for several hundred million objects over $\sim$33000 deg$^{2}$ between $-75^{\circ} < {\mathrm Dec} < 75^{\circ}$ (except for within $\sim$10 -- 15$^{\circ}$ of the Galactic plane) to $\sim$20 mag. and with an average of $\sim$250 observations over a 7-year baseline. $\sim$200 million light curves from the CRTS DR1 are already accessible from both the CRTS web site\footnote{http://nesssi.cacr.caltech.edu/DataRelease} and the VAO Time Series Search Tool\footnote{http://www.usvao.org/science-tools-services/time-series-search-tool}. The latter site (see Fig.~\ref{fig1}) is a pathfinder utility for interconnecting repositories of time series data. and provides access to important data sets at the NASA Exoplanet Archive\footnote{http://exoplanetarchive.ipac.caltech.edu}, such as Kepler, CoRoT, HATNet, TrES and KELT,  and at the Harvard Time Series Center\footnote{http://timemachine.iic.harvard.edu/}, such as ASAS, in addition to the CRTS DR1 data. The service also offers access to some analysis tools (see below). 

\section{Event and source cross-identification}

A component of the VOEvent architecture\cite{voevent} is an {\em annotator}, which is a service that receives an event notification and acts on it in such a way that a subsequent event is generated adding information to what is already known about the celestial event that triggered the original event notification.
One of the most common annotation activities is to cross-identify events with other data archives, i.e., search for plausible spatial associations between an event and other observations, typically at other wavelengths (the VAO Cross-Comparison Tool\footnote{http://www.usvao.org/science-tools-services/cross-comparison-tool} performs fast positional cross-matches between an input table of up to 1 million sources and common astronomical source catalogs, such as SDSS and 2MASS). The simplest matching criterion is just to take the nearest positional hit but this is not necessarily the best match. Positional accuracies can vary widely between surveys, particularly between different wavelength regimes, leading to multiple possible crossmatch candidates, e.g., a brighter object might have a smaller positional error due to its stronger detection whereas a fainter object might be farther away yet still as likely due to its larger positional error. Other information may also make certain matches far more likely, such as a potential supernova being more likely associated with a nearby galaxy than a star. Several formalisms have been proposed to deal with the general problem of spatial crossmatching but Budavari \cite{budavari} uses Bayesian hypothesis testing to evaluate the quality of candidate associations specifically for detections in space {\em and} time thus allowing inclusion of information about the temporal behavior of particular sources.

\begin{figure}
\includegraphics[width=6.78in]{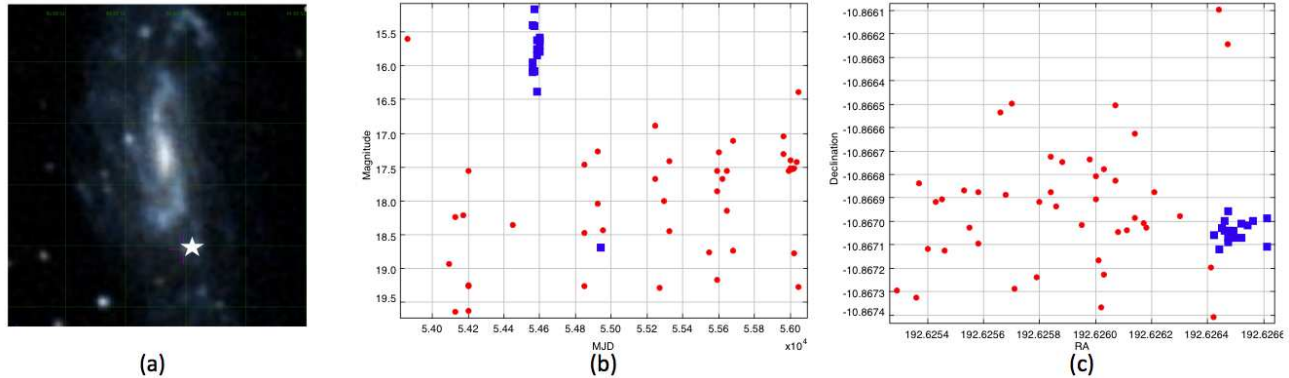}
\caption{This shows how positional cross-matching information can be used to correctly identify variable sources. The full signal from the supernova 2008aq (a) in the CRTS  light curve (b) (squares) only becomes apparent when the spatial positions are examined (c).}
\label{fig2}       % Give a unique label
\end{figure}

A related activity is constructing the time histories of astronomical objects from sets of individual observations of them within the same survey. Fig.~\ref{fig2}(b) shows the CRTS time series for the position associated with SN 2008aq. The light curve for the supernova (squares) is contaminated by observations of nearby sources (points) since it occurred in the outer reaches of a spiral galaxy (MCG-02-33-20, Fig.~\ref{fig2}(a)). Only by crossmatching the positions of individual observations can the subgroup that corresponds to the actual supernova be identified; in particular, the late data point from the supernova when it has faded by $\sim$3 mags. which would otherwise be lost in the background signal from the galaxy. The Budavari formalism can recover this information on the assumption of a suitable prior model for the supernova light curve. 

Recovering the light curve for a single source is generally a straightforward operation but for every object in a survey, it soon becomes intractable. Constructing the (full transitive) set of associations for $n$ sources from a set of $m$ observations scales at best as ${\cal O}(\sim\!nm^2)$, assuming only one match between individual sets (note that ${\cal O}(\sim\!n^2)$ is normally already considered prohibitive in many circles). The Palomar-Quest DR1 (PQDR1) catalog of $\sim$10 million sources with typically $\sim$15 observations per source resulted in a set of over 4 billion associations; with $\sim$500 million sources, each with typically $\sim$200 observations per source, CRTS would have at least $\sim$20 trillion associations. The next generation surveys will take us into the quadrillions and beyond. These are also inherently probabilistic groupings. Varying conditions between observations -- sky brightness, atmospheric, instrumental, etc. -- mean that the same detection thresholds and positional errors cannot be assumed across a survey, e.g., perfect conditions may mean that two nearby sources are resolved on one night but appear as a blended source on another poorer quality night so a single observation may well be associated with multiple histories, particularly in crowded fields.

An obvious scalable solution to maintain a master catalog and update associations on a per night basis (using a service such as the VAO Cross-Comparison Tool) rather than associate all data in a single go. However, this latter operation will still become necessary when the master catalog needs to be revised, e.g., with improved positional error models. It is interesting to note that whilst spatial pixellation schemes, which provide a single identifier for a region of sky, such as HTM \cite{htm} and Healpix \cite{healpix}, are good for individual object lookups in catalogs, there are more efficient ways of doing bulk crossmatches. 
The Zones algorithm \cite{zones} developed for the SDSS and 2MASS surveys uses a B-tree to bucket two-dimensional space giving dynamically computed bounding boxes (B-tree ranges) for spatial queries. In practice, using an optimal zoning gives several factors of ten increased performance over using indexing schemes, although, in tests with PQDR1, the Quad Tree Cube scheme \cite{q3c} has also shown itself to be equivalently fast.

\section{Source characterization}
When a significant variation in an astronomical source is detected (the significance is determined by such factors as the size, suddenness and duration of the variation as well as the type of detector used), an event notification (VOEvent) is broadcast to all interested parties. This triggers a cascade of activity where the event is placed in context with related data and information: followup observations of the same astronomical event (if possible), source cross-identifications, etc. This {\em data portfolio} for the event represents a summation of all that is known and understood about it. The most interesting or exciting events will be associated with rich portfolios containing a wide range of heterogeneous material whereas a commonplace event might have a portfolio containing only the original event notification. Note that a portfolio is also a dynamic entity with the potential for new material to be added at any time, from milliseconds to years or even decades after the initial event. 

For any analysis activity involving the comparison of events (or sources), the heterogeneity of information has to be replaced by a common characterization in terms of a representative set of features.  Even when dealing just with light curves, there can be tremendous disparity between temporal coverage, sampling rates and regularity, number of points and error bars, e.g., a transient light curve might consist of a couple of points and a lot of upper limits (when the source was below the detection threshold of the survey) whilst a monitored exoplanet candidate light curve may contain hundred of thousands of high signal-to-noise points. A number of recent papers \cite{richards, dubath, shin} have explored a variety of characterizing features for the light curves of known variable stars, including statistical moments, flux and shape ratios, variability indices, periodicity measures, and model representations. In this vein, the Caltech Time Series Characterization Service\footnote{http://nirgun.caltech.edu:8000}  (CTSCS) is an experimental service which aims to extract a comprehensive set of (over 60) such features from any supplied light curve. 

Many features employed in characterizing light curves are founded on some type of periodic analysis of the time history of the object. Unfortunately there is no single way to determine the period of a light curve that is accurate and reliable - for example, the Lomb-Scargle method, which is in many ways the de facto standard technique, is at best $\sim$75\% accurate and less so with irregular sampling strategies and light curves with small numbers of points\cite{graham}. The NASA Exoplanet Archive Periodogram Service\footnote[1]{http://exoplanetarchive.ipac.caltech.edu/cgi-bin/Periodogram/nph-simpleupload} supports three algorithms: Lomb-Scargle, box-fitting least squares and Plavchan, as well as control over their various parameters so that they can be fine-tuned for a particular type of time series data. An advantage of this service is that it is integrated with the VAO Time Series Search Tool so a period can be easily obtained for a discovered light curve without having to download the data first and then upload it to the service. The CTSCS offers more period-finding algorithms but is not integrated with the Time Series Tool. The NASA tool is also designed around an enterprise architecture and so can handle larger workloads.

\section{Source classification}
Source classification is a very challenging problem, particularly for transient events, due to the sparsity and heterogeneity of the available data. Some current efforts on the classification of transients in the optical synoptic sky surveys, employing both the source light curve and its data portfolio, include Refs.~\citenum{bloom, donalek, mahabal1, mahabal2, mahabal3, mahabal4, djorgovski}. The Caltech-JPL group, in particular, has experimented with several approaches, using data from the Palomar-Quest survey\cite{pq} and Catalina Real-time Transient Survey\cite{crts, mahabal4, drake11}, as well as selected data sets from the literature.

When addressing archival (i.e., not real time) classification, light curves are reduced to common basis feature vectors (as described above) as input to machine learning techniques (e.g., decision trees (DT)).
In particular, DTs have been trained using the feature vectors for various combinations of known classes of object. To reduce the dimensionality of the input space, a forward feature selection strategy is applied that selects the best subset of features from a training data set that can predict test data. The DTs themselves are built using the Gini diversity index as the splitting criterion with 10-fold cross validation to avoid overfitting. Moreover, the DTs in the iterations are pruned in order to choose the simplest one within one standard error of the minimum. When tested on a data set consisting of light curves of blazars, cataclysmic variables, and RR Lyrae stars, the method\cite{donalek12} achieves between $\sim$ 83\% and 97\% completeness and $\sim$ 4\% to 13\% contamination. This approach seems very promising for the classification of radio light curves as well.

Another novel approach\cite{djorgovski11, moghaddam} uses 2-dimensional distributions of magnitude changes for different time baselines for all possible epoch pairs in the data set.  These 2-dimensional ($\Delta m, \Delta t$) histograms can be viewed as probabilistic structure functions for the light curves of different types.  Template distributions for different kinds of transients and variables are constructed using the reliably-classified data with the same survey cadences, S/N, etc.  For any newly detected variable or a transient, corresponding ($\Delta m, \Delta t$) histograms are accumulated as the new data arrive, and a variety of metrics used to compute the effective probabilistic distances from different templates.  The tests so far indicate that classification accuracies in excess of 90\% may be possible using this approach.  Generalizations to include triplets or even higher order sets of data points for multi-dimensional histograms are planned.

Other approaches have been tested as well but their description is beyond the scope of this paper.
However, it quickly became very clear that different classifiers may be optimal for different types of transient or variable sources. Thus, there is a need for a meta-classifier that would provide an optimal classification for any given event on the basis of several different classifier results.  This work is still in progress.

One important lesson learnt so far is that the existing archival and contextual data will play a critical role in classification.  This includes a spatial context (i.e., what is near the observed event on the sky - a possible host galaxy, a cluster, a SN remnant, etc.), the multiwavelength context (has it been detected on other wavelength, what is its broad spectral energy distribution), and the temporal context (what was its flux variability or a detection history in the previously obtained data). In fact, most of the relevant information in hand when a transient is first detected is of this nature.  Some of it can be readily extracted from the archives using VAO tools (e.g., flux measurements form different wavelengths, light curves at that location), but some - spatial context in particular - require a human judgment, e.g., is the apparent proximity to other objects (galaxies, clusters, etc.) likely to be relevant, and if so, what does it imply about the transient? Human inspection of vast numbers of transients does not scale to the massive data streams such as those contemplated here.  Crowdsourcing approaches to harvesting the relevant human pattern recognition skills and domain expertise, and their translation to machine-processable algorithms are being investigated.

It is intended that automated transient source classifiers for CRTS detections will be deployed as annotator services in the near future. Classifications for individual events will be broadcast as followup VOEvents and employ a machine-processible concept scheme, e.g., the IVOA thesaurus or Unified Astronomy Thesaurus, to describe the most likely object type. This will allow robotic telescopes to trigger followup observations not only on parameter-based rules, e.g., magnitude $< 17$, but also on (probabilistic) classifications described with community-standard terms.

\section{Conclusions}
The emerging field of time domain astronomy requires tools and infrastructure to support a distributed network of (massive) real-time data streams, data archives, and analysis services. The VAO is developing an interoperable framework to connect partner providers of both data and analysis resources, and expose them as an integrated whole for wider community use. A recent community-wide call for collaborative proposals by the VAO\cite{vao} has identified two time domain projects which it is now advising. One is concerned with access to data related to the Variable and Slow Transient (VAST) Survey Science Program of the Australian Square Kilometre Array Pathfinder (PI: T. Murphy) and the other involves access to the databases of the American Association of Variable Star Observers (AAVSO, PI: M. Templeton). Such collaborations, combining domain expertise in data technologies and the relevant science areas, illustrate the potential of 
an informatics-based approach to data-intensive science.

\acknowledgments     %>>>> equivalent to \section*{ACKNOWLEDGMENTS}      
Support for the development of time series infrastructure is provided by the Virtual Astronomical Observatory contract AST-0834235. Work on source characterization and classification has been supported in part by the National Science Foundation grants AST-0407448, CNS-0540369, AST-0909182 and IIS-1118041; the National Aeronautics and Space Administration grant 08-AISR08-0085; and by the Ajax and Fishbein Family Foundations. We are thankful to numerous colleagues in the VO and Astroinformatics community, and to the members of the DPOSS, PQ, and CRTS survey teams, for many useful discussions and interactions through the years. 

%%%%%%%%%%%%%%%%%%%%%%%%%%%%%%%%%%%%%%%%%%%%%%%%%%%%%%%%%%%%%
%%%%% References %%%%%

% Responding to the event deluge, Roy D. Williams, California Institute of Technology (USA); Scott Barthelmy, NASA Goddard Space Flight Ctr. (USA) . . . . . . . . . . . . . . . . . . . . . . . . . . . . . . . . . . . . . . . . . . . . . . . . . . . . [8448-26]
\bibliography{paper}   %>>>> bibliography data in report.bib
\bibliographystyle{spiebib}   %>>>> makes bibtex use spiebib.bst

\end{document}